\documentclass[twocolumn,floatfix,english,superscriptaddress,citeautoscript]{revtex4-2}
\usepackage[T1]{fontenc}
\usepackage[latin9]{inputenc}
\usepackage{array}
\usepackage{multirow}
\usepackage{amsmath}
\usepackage{amssymb}
\usepackage{graphicx}
\usepackage{placeins}
\usepackage{esint}
\usepackage{xcolor}
\newcommand{\dbar}{\mathchar'26\mkern-12mu d}

\makeatletter

\providecommand{\tabularnewline}{\\}

\usepackage{wasysym}

\providecommand{\tabularnewline}{\\}

\usepackage{babel}

\makeatother

\usepackage{babel}
\usepackage[colorlinks=true,linkcolor=blue,citecolor=blue,urlcolor=blue]{hyperref}
\begin{document}
\title{A single-electron double quantum dot with Rashba spin-orbit interaction as a working substance for heat machines}
\author{Ivan Romualdo de Oliveira}
\affiliation{Department of Physics, Institute of Natural Science, Federal University
of Lavras, 37200-900 Lavras-MG, Brazil}
\author{Vin\'icius N. A. Lula-Rocha}
\affiliation{Department of Physics, Institute of Natural Science, Federal University
of Lavras, 37200-900 Lavras-MG, Brazil}
\author{Moises Rojas}
\affiliation{Department of Physics, Institute of Natural Science, Federal University
of Lavras, 37200-900 Lavras-MG, Brazil}

\begin{abstract}
We investigate the thermodynamic performance of a quantum Otto machine whose working substance is a single electron confined in a double quantum dot under an external magnetic field and Rashba spin-orbit interaction. The Hamiltonian is controlled by the Zeeman splitting, the interdot tunneling amplitude, and the Rashba coupling, which induces spin-flip tunneling between localized orbital states. Within a quasistatic Otto cycle, we analyze the heat exchanged with the reservoirs, the extracted work, and the efficiency as functions of the Hamiltonian parameters and reservoir temperatures. We show that the Rashba interaction acts as an effective control parameter for switching among heat-engine, refrigerator, heater, and accelerator regimes. A global numerical analysis over the Hamiltonian parameters and reservoir temperatures identifies the optimal operating points for efficiency and work output in the heat-engine regime. The highest efficiencies occur near the maximum temperature gradient explored and approach the Carnot bound, whereas the largest work output appears in a different region of parameter space. The results reveal a clear trade-off between maximum efficiency and maximum extracted work, governed by the spectral deformation induced by the Zeeman splitting, tunneling amplitude, and Rashba coupling.
\end{abstract}
\maketitle

\section{Introduction}

Quantum thermodynamics is an emerging field that investigates the interplay between thermodynamic principles and quantum mechanics, particularly in systems where quantum features play a central role \cite{deff,binder,can}. It explores how inherently quantum phenomena---such as coherence, entanglement, and nonequilibrium dynamics---affect thermodynamic processes and the performance of quantum thermal machines \cite{kos,alicki,jo,pet}. In this context, the working medium can range from spin systems \cite{su,sod,abd,moi} and harmonic oscillators \cite{in,kos-1} to multi-level quantum systems \cite{moi-1}, ideal quantum gases \cite{pathria}, nanomechanical heat engine \cite{jit}, superconducting spintronic heat engine \cite{clo}, mesoscopic quantum heat \cite{pat}, and even quantum dots \cite{fran,clev}. These systems exhibit behaviors that deviate markedly from classical expectations, offering novel mechanisms and the potential for enhanced efficiency. Recent advances have demonstrated that quantum engines and refrigerators can exploit features such as energy-basis coherence, ergotropy, and non-thermal reservoirs to outperform their classical counterparts \cite{vin,sid,das,cam,vinicius}. These developments not only address practical challenges in nanoscale engineering but also deepen our understanding of the fundamental links between statistical mechanics, thermodynamics, and quantum information science \cite{binder,can}.

Scalable platforms for quantum information processing have motivated intense interest in quantum dot (QD)--based systems, in which confined electrons act as qubits through their spin or charge degrees of freedom \cite{loss,shin,petta,bur,moi-2,zak}.  Moreover, the quantum dynamics and entanglement of two electrons in coupled double quantum dots (DQDs) have been extensively  investigated \cite{kess,fan,bra}, together with studies of quantum correlations and decoherence effects \cite{cao,ben,souza}. Related works have addressed microwave control of two-electron spin states \cite{cer}, leading the enhanced coherence times and gate fidelities \cite{tay,nic}, as well as, quantum teleportation protocols based on DQDs \cite{rao,lane,zopf}.

Beyond their role as qubit platforms, DQDs also provide a versatile solid-state architecture for quantum thermal machines, where gate voltages and tunnel couplings enable \textit{in situ} control of level detuning, hybridization, and energy filtering. In capacitively coupled DQDs, heat-to-work conversion can be realized in three-terminal geometries that spatially separate the heat source from the charge circuit, allowing thermally induced charge currents to be generated by thermal fluctuations and tuned by external gates \cite{sanchez,thierschmann,sothmann}. More recently, serial DQD devices have been explored as nanoscale heat engines in which phonon-assisted and thermoelectric transport mechanisms can be disentangled and optimized by tuning the interdot coupling \cite{dorsch}. Related studies have also explored systems composed of two coupled double quantum-dot systems operating as the working substance of quantum heat machines \cite{clev}. Spin--orbit interaction (SOI) in semiconductor nanostructures arises from the absence of inversion symmetry, generating effective momentum-dependent magnetic fields that couple orbital motion to the spin degree of freedom. In zinc-blende materials, SOI typically includes both Rashba (structural inversion asymmetry) and Dresselhaus (bulk inversion asymmetry) contributions \cite{rashba,dresselhaus,hanson}. In gate-defined quantum dots, SOI enables all-electric spin manipulation via electric-dipole spin resonance and, in double quantum dots, it can mediate spin-flip tunneling between the two localized orbitals \cite{golovach,nowack}. The relevance of SOI in semiconductor quantum-dot systems has therefore attracted considerable attention, particularly due to its potential for coherent spin-control and orbit-assisted quantum memories based on double quantum dots \cite{cho}, as well as the investigation by Li et al. of Rashba-induced effects on quantum gate operations with simultaneous transport in double quantum dots \cite{muga}. More recently, the thermal entanglement and quantum coherence of a single electron in a double quantum dot with Rashba interaction have also been investigated \cite{mrojas-1}.

Quantum heat engines constitute a promising frontier where thermodynamics and quantum mechanics intersect. In this work, we investigate a minimal yet physically rich realization of a quantum Otto cycle based on a single electron confined in a double quantum dot, simultaneously subject to an external magnetic field and Rashba spin-orbit interaction \cite{mrojas-1}. Operating in the strong Coulomb-blockade regime, with exactly one electron occupying the DQD, we adopt a quasistatic framework in which, at the end of each isochoric branch of the thermodynamic cycle, the system is described by a Gibbs state, thereby ensuring thermal equilibrium with the corresponding reservoir. In this setting, the electron mediates the energy exchange, while orbital tunneling and spin-flip transitions induced by spin-orbit coupling play a central role in the system dynamics. Within this framework, we address a fundamental question: how do quantum correlations arising from the competition among Zeeman splitting, interdot tunneling, and spin-orbit interaction affect the thermodynamic performance of the quantum heat engine?

This paper is organized as follows. In Sec. \ref{theModel} we present the double quantum dot system that plays the role of the working substance. In Sec. \ref{quantumThermodynamics} we review the main aspects of quantum thermodynamics relevant to this work. We present the theory of the quantum Otto cycle applied to our model in Sec. \ref{quantumOttoCycle}, and in Sec. \ref{results} we analyze the thermodynamic behavior of the double quantum dot for different reservoir temperatures and Hamiltonian parameters. Finally, Sec. \ref{conclusionsAndPerspectives} is devoted to our conclusions and perspectives.

\section{The model}
\label{theModel}

We consider a single-electron double quantum dot modeled as a double potential well under an external magnetic field and Rashba spin-orbit coupling.

The state of the system is controlled by three real parameters: $\Delta$, $t$ and $\alpha$. The parameter $\Delta$ controls the Zeeman coupling between the magnetic field and the electron spin, whose two states are represented by the basis states $\{|0\rangle, |1\rangle\}$. The parameter $t$ controls the probability amplitude for the electron to tunnel between the left ($L$) and right ($R$) wells, associated with the states $\{|L\rangle, |R\rangle\}$. Finally, $\alpha$ is the coupling constant of the Rashba spin-orbit interaction. This interaction has the interesting property of flipping the spin state while the electron is exchanged between the two potential wells. Thus, the Hilbert space of the system is spanned by the basis $\{ |L0\rangle, |L1\rangle, |R0\rangle, |R1\rangle \}$ and it is described by the Hamiltonian
\begin{eqnarray}
H
&=&
\frac{\Delta}{2}
\left(
\mathbb{I}\otimes \sigma_{z}
\right)
+
t
\left(
\tau_{x}\otimes \mathbb{I}
\right)
+
\alpha
\left(
\tau_{y}\otimes \sigma_{x}
\right),
\label{Model1}
\end{eqnarray}
where
\begin{eqnarray}
\sigma_{x}
&=&
|0\rangle\langle 1| + |1\rangle\langle 0|,
\label{Model2}
\\
\sigma_{z}
&=&
|0\rangle\langle 0| - |1\rangle\langle 1|,
\label{Model3}
\\
\tau_{x}
&=&
|L\rangle\langle R| + |R\rangle\langle L|,
\label{Model4}
\\
\tau_{y}
&=&
-i|L\rangle\langle R| + i|R\rangle\langle L|.
\label{Model5}
\end{eqnarray}
The Hamiltonian eigenvector is given by
\begin{eqnarray}
\!\!\!\!
|\varphi_{\pm}\rangle
&=&
A_{\pm}
\left[
i a_{\pm}
\left(
|L0\rangle + |R0\rangle
\right)
- 
\left(
|L1\rangle - |R1\rangle
\right)
\right],
\label{Model7}
\\
|\Phi_{\pm}\rangle
&=&
B_{\pm}
\left[
i b_{\pm}
\left(
|L0\rangle - |R0\rangle
\right)
+ 
\left(
|L1\rangle + |R1\rangle
\right)
\right],
\label{Model7b}
\end{eqnarray}
where 
$
A_{\pm}
= 
\frac{1}{
	\sqrt{2}
	\sqrt{
		a_{\pm}^{2} + 1
	}
}
$,
$
a_{\pm}
=
\frac{
	\Omega_{+} 
	\pm
	\sqrt{
		\Omega_{+}^{2} + 4\alpha^{2}
	}
}{2\alpha}
$,
$
B_{\pm}
=
\frac{1}{
	\sqrt{2}
	\sqrt{b_{\pm}^{2} + 1}
}
$,
$
b_{\pm}
=
\frac{
	\Omega_{-}
	\pm
	\sqrt{
		\Omega_{-}^{2} + 4\alpha^{2}
	}
}{2\alpha}
$
and
$\Omega_{\pm} = \Delta \pm 2t$. The corresponding eigenvalues are
\begin{eqnarray}
E_{\varphi_{\pm}}
&=&
\pm
\frac{1}{2}
\sqrt{
	\Omega_{+}^{2} + 4\alpha^{2}
},
\label{Model8}
\\
E_{\Phi_{\pm}}
&=&
\pm
\frac{1}{2}
\sqrt{
	\Omega_{-}^{2} + 4\alpha^{2}
}.
\label{Model9}
\end{eqnarray}
We are interested in the system's state when it is subjected to thermal equilibrium with thermal reservoirs at inverse of temperature $\beta=1/k_{B}T$, where $k_{B}$ is the Boltzmann constant and $T$ is the reservoir's absolute temperature. This is given by the Gibbs state
\begin{eqnarray}
\rho(\beta,H)
&=&
\frac{
	e^{-\beta H}
}{Z(\beta)},
\label{Model10}
\end{eqnarray}
where $Z(\beta)=\textrm{Tr}[e^{-\beta H}]$ is the partition function and $H$ is the Hamiltonian (\ref{Model1}).

\section{Quantum thermodynamics}
\label{quantumThermodynamics}

The key ingredients of quantum thermodynamics relevant to our work are the following. From the first law of classical thermodynamics, the variation of the internal energy $dE$ of a system is equal to the heat absorbed by the system minus the work performed by the system,
\begin{eqnarray}
dE
&=&
\dbar Q_{cl} - \dbar W_{cl}.
\label{QuantumThermodynamics1}
\end{eqnarray}
At the microscopic scale, there are two complementary ways of understanding internal energy. The first comes from the statistical-mechanical description of the canonical ensemble, which states that the average internal energy $E$ of a system governed by a Hamiltonian $\mathcal{H}$ is given by
\begin{eqnarray}
E
&=&
\textrm{Tr}
\left[
\hat{\rho}\hat{\mathcal{H}}
\right],
\label{QuantumThermodynamics2}
\end{eqnarray}
where $\hat{\rho}$ is the density operator that describes the state of the quantum system whose energy is represented by the Hamiltonian operator $\hat{\mathcal{H}}$.

The second is the quantum-mechanical understanding of the energy average as the expectation value of the energy observable represented by $\hat{\mathcal{H}}$ in a state $|\psi\rangle$, given by $\langle \hat{\mathcal{H}} \rangle = \langle \psi| \hat{\mathcal{H}}|\psi\rangle$.  

In quantum thermodynamics these concepts are understood to be the same quantity. For this reason we are allowed to write
\begin{eqnarray}
E
&=&
\langle \hat{\mathcal{H}} \rangle
=
\textrm{Tr}
\left[
\hat{\rho}\hat{\mathcal{H}}
\right].
\label{QuantumThermodynamics3}
\end{eqnarray}
Thus, the variation of the internal energy of a quantum system can be described as
\begin{eqnarray}
dE
=
d\langle \hat{H} \rangle
&=&
\textrm{Tr}[d\hat{\rho}\hat{\mathcal{H}}] 
+
\textrm{Tr}[\hat{\rho} d\hat{\mathcal{H}}].
\label{QuantumThermodynamics4}
\end{eqnarray}
The second term on the r.h.s. of Eq.~(\ref{QuantumThermodynamics4}) represents the variation of the Hamiltonian while the state of the system remains unchanged. This means that the parameters of the operator describing the energy of the system, such as the magnetic field, change while the probabilities of occupying the energy levels do not. Changing the Hamiltonian parameters modifies the energy-level gaps while preserving their populations. In a classical analogy, one can say that the system produces or receives work as in a compression or expansion process. Here, compression and expansion refer to changes in the energy gaps, in analogy with the compression and expansion of a gas in classical thermodynamics. Since we adopt the convention that positive work is work performed by the system, the quantum concept of work $\dbar W_{quan}$ is given by
\begin{eqnarray}
\dbar W_{quan}
&=&
-\textrm{Tr}
\left[
\hat{\rho}d\hat{\mathcal{H}}
\right].
\label{QuantumThermodynamics5}
\end{eqnarray}
The first term of Eq.~(\ref{QuantumThermodynamics4}) represents the variation of the quantum state itself while the Hamiltonian remains the same. In this case, the contribution to the energy average comes only from the internal state configuration. In other words, the energy gaps remain fixed while the probabilities of occupying these levels change. This is the direct analogue of the classical notion of heat. For example, consider a box with constant volume containing a gas. A flux of heat transferred to the gas increases the kinetic energy of the particles and, consequently, the internal energy. Since the box has constant volume, no work is produced. Therefore, we can identify quantum heat $\dbar Q_{quan}$ as
\begin{eqnarray}
\dbar Q_{quan}
&=&
\textrm{Tr}
\left[
\hat{\mathcal{H}} d\hat{\rho}
\right].
\label{QuantumThermodynamics6}
\end{eqnarray}
Performing the trace operation in energy eigenbasis $\{|n\rangle\}$, $\hat{\mathcal{H}}|n\rangle = E_{n}|n\rangle$, with the assumption that $|n\rangle$ does not vary with time, we arrive in suitable expressions for (\ref{QuantumThermodynamics5}) and (\ref{QuantumThermodynamics6}), given by
\begin{eqnarray}
\dbar W_{quan}
&=&
-\sum_{n} 
P_{n}dE_{n},
\label{QuantumThermodynamics7}
\\
\dbar Q_{quan}
&=&
\sum_{n}
E_{n}dP_{n},
\label{QuantumThermodynamics8}
\end{eqnarray}
where $P_{n} = \langle n|\hat{\rho}|n\rangle$. From here on $W$ and $Q$ symbols and the respective words work and heat will always denote the notion of quantum work and quantum heat.

\section{Quantum Otto cycle}
\label{quantumOttoCycle}

The quantum Otto cycle is the quantum counterpart of the classical Otto cycle. It is comprised of a working substance, two thermal reservoirs at different temperatures, and four stages: two quasistatic adiabatic (unitary) processes and two isochoric thermalization strokes, in which the working substance exchanges heat with the reservoirs.

Our working substance is the double quantum dot in an external magnetic field with Rashba interaction, as described in Sec.~\ref{theModel}. We assume that the working substance and the two thermal reservoirs form an isolated composite system, in the sense that no energy is exchanged with the external environment. The Hamiltonian parameters $\Delta$, $t$, and $\alpha$ can be tuned externally.

To clarify the notation, we consider two thermal reservoirs: a cold reservoir with inverse temperature $\beta_{c}$ and a hot reservoir with inverse temperature $\beta_{h}$. The Hamiltonian parameters will be denoted by the cold set $S_{c}=\{\Delta_{c},t_{c},\alpha_{c}\}$ and the hot set $S_{h}=\{\Delta_{h},t_{h},\alpha_{h}\}$, and we write $H_{c}=H(S_{c})$ and $H_{h}=H(S_{h})$. The working substance in thermal equilibrium with the cold and hot reservoirs will be denoted by $\rho_{c}=\rho(\beta_{c},H_{c})$ and $\rho_{h}=\rho(\beta_{h},H_{h})$, respectively, where $\rho(\beta,H)$ is the Gibbs state in Eq.~(\ref{Model10}).

We denote $\rho_{h}(H_{c}) = U\rho_{h}U^{\dagger}$ where $U$ is a unitary transformation that relates the eigenvectors $\{|n_{c}\rangle\}$ of $H_{c}$ and the eigenvectors $\{|n_{h}\rangle\}$ of $H_{h}$, $|n_{c}\rangle= U|n_{h}\rangle$. As eigenvectors of Hamiltonians, we have that $H_{c}|n_{c}\rangle = E_{n}^{c}|n_{c}\rangle$ and $H_{h}|n_{h}\rangle = E_{n}^{h}|n_{h}\rangle$, where $E_{n}^{c}$ and $E_{n}^{h}$ are their respective eigenvalues. We also have that $\rho_{c}(H_{h}) = U^{\dagger}\rho_{c}U$. The variation of the density operator $d\rho$ will be considered the difference between the final $\rho_{f}$ and initial $\rho_{i}$ states in consideration in their respective processes, $d \rho = \rho_{f} - \rho_{i}$.

Since $\rho_{c}$ and $\rho_{h}$ are Gibbs states with respect to $(\beta_{c},H_{c})$ and $(\beta_{h},H_{h})$, respectively, they are diagonal in the eigenbases of their corresponding Hamiltonians. Thus,
$\rho_{c}|n_{c}\rangle=p_{n}^{c}|n_{c}\rangle$ and $\rho_{h}|n_{h}\rangle=p_{n}^{h}|n_{h}\rangle$, where $p_{n}^{c}$ and $p_{n}^{h}$ are eigenvalues (populations). In addition, we define $p_{n}^{h}(S_{c})=\langle n_{c}|\rho_{h}(H_{c})|n_{c}\rangle$ and $p_{n}^{c}(S_{h})=\langle n_{h}|\rho_{c}(H_{h})|n_{h}\rangle$.

Now we detail the four stages of the quantum Otto cycle.

{\bf 1. } We begin with the working substance in thermal equilibrium with the hot reservoir, so it is represented by the state $\rho_{h}$ and the Hamiltonian is $H_{h}$. We then decouple the working substance from the reservoir and perform the unitary evolution on $\rho_{h}$, which consists of changing the Hamiltonian parameters $H_{h} \mapsto H_{c}$ while the populations in the instantaneous energy eigenbasis remain the same, that is $p_{n}^{h}(S_{c})=\langle n_{c}|\rho_{h}(H_{c})|n_{c}\rangle =\langle n_{h}| \rho_{h}|n_{h}\rangle = p_{n}^{h}$. This process is quasistatic and adiabatic and is given by $\rho_{h}(H_{c})=U\rho_{h}U^{\dagger}$. Since the populations are adiabatically transported, $dP_n=0$ in the instantaneous energy eigenbasis, and therefore Eq.~(\ref{QuantumThermodynamics8}) gives no heat exchange in this stroke. The variation of the average energy is completely due to work associated with the parameter change. The work in this adiabatic stroke is denoted by $W_{c}$, where the subscript indicates the final cold configuration, and is given by
\begin{eqnarray}
W_{c}
&=&
\sum_{n}
p_{n}^{h}
\left(
E_{n}^{h} - E_{n}^{c}
\right),
\label{OttosCycle1}
\end{eqnarray}
where $n$  runs from $1$ to $4$, where each $n$ corresponds to one element of the set of energy subindices $\{\varphi_{+}, \varphi_{-}, \Phi_{+}, \Phi_{-}\}$ in equations   (\ref{Model8}) and (\ref{Model9}).

{\bf 2.} In this stage, the initial state of the working substance is $\rho_{h}(H_{c})$ and it is coupled to the cold reservoir until it is quasistatically thermalized at inverse temperature $\beta_{c}$, i.e., $\rho_{h}(H_{c})\mapsto\rho_{c}$. In this process, the Hamiltonian is constant ($H_{c}$), so the change in the average energy is entirely due to the change in the density operator. Consequently, no work is performed and the working substance exchanges heat $Q_{c}$ with the cold reservoir, given by
\begin{eqnarray}
Q_{c}
&=&
\sum_{n}
E_{n}^{c}
\left(
p_{n}^{c} - p_{n}^{h}(S_{c})
\right).
\label{OttosCycle2}
\end{eqnarray}
Since the unitary stroke preserves populations, we have $p_{n}^{h}(S_{c})=p_{n}^{h}$, and the heat in this stage can be written as
\begin{eqnarray}
Q_{c}
&=&
\sum_{n}
E_{n}^{c}
\left(
p_{n}^{c} - p_{n}^{h}
\right).
\label{OttosCycle3}
\end{eqnarray}

{\bf 3.} This process begins with the working substance in thermal equilibrium with the cold reservoir, so its state is given by $\rho_{c}$. We then decouple it from the reservoir. This stage is analogous to process {\bf 1.}, but the quasistatic adiabatic process now reads $\rho_{c} \mapsto \rho_{c}(H_{h}) = U^{\dagger}\rho_{c}U$. The unitary transformation ensures that the populations remain the same at the end of the process, since $p_{n}^{c}(S_{h}) = \langle n_{h}|\rho_{c}(H_{h})|n_{h}\rangle = \langle n_{c}| \rho_{c}|n_{c}\rangle = p_{n}^{c}$. As in the first adiabatic stroke, the populations are preserved in the instantaneous energy eigenbasis, so $dP_n=0$ and no heat is exchanged with the environment. The work in this adiabatic stroke is denoted by $W_{h}$, where the subscript indicates the final hot configuration, and is given by
\begin{eqnarray}
W_{h}
&=&
\sum_{n}
p_{n}^{c}
\left(
E_{n}^{c} - E_{n}^{h} 
\right).
\label{OttosCycle4}
\end{eqnarray}

{\bf 4.} In this process, the working substance begins in the state $\rho_c(H_{h})$ and is coupled to the hot reservoir. The state evolves until it reaches thermal equilibrium with the reservoir at inverse temperature $\beta_{h}$, so $\rho_c(H_{h})\mapsto \rho_{h}$. As in {\bf 2.}, the Hamiltonian remains constant, in this case $H_{h}$, so no work is performed and the energy variation is entirely due to the density-operator variation $d\rho_{4} = \rho_{h} - \rho_{c}(H_{h})$. The working substance exchanges heat $Q_{h}$ with the hot reservoir, given by
\begin{eqnarray}
Q_{h}
&=&
\sum_{n}
E_{n}^{h}
\left(
p_{n}^{h} - p_{n}^{c}(S_{h})
\right).
\label{OttosCycle5}
\end{eqnarray}
Since the unitary transformation ensures that $p_{n}^{c}(S_{h}) = p_{n}^{c}$, Eq.~(\ref{OttosCycle5}) can be written as
\begin{eqnarray}
Q_{h}
&=&
\sum_{n}
E_{n}^{h}
\left(
p_{n}^{h} - p_{n}^{c} 
\right).
\label{OttosCycle6}
\end{eqnarray}
This process ends with the working substance in the state $\rho_{h}$, which will be the initial state of process {\bf 1.}, closing the cycle.

The total work $W_{T}$ in the Otto cycle can be calculated by $W_{T} = W_{c} + W_{h}$. Another way of calculating the total work is by noting that the variation of the total energy in a closed cycle is null, so by the first law of thermodynamics,
\begin{eqnarray}
W_{T}
&=&
Q_{T},
\label{OttosCycle7}
\end{eqnarray}
where $Q_{T} = Q_{c} + Q_{h}$ is the total heat.
In the following, we denote the total work simply by $W\equiv W_T$.


\section{Results}
\label{results}

In this section we analyze the regime of operation and thermodynamic performance of the quantum Otto cycle by focusing on the influence of the Rashba spin-orbit interaction, which is implemented in our model via the parameter $\alpha$. In particular, we are interested in understanding how the change of this parameter affects the thermodynamic regime of the machine. Since the Rashba term couples the spin degree of freedom to the tunneling between the two dots, variations of $\alpha$ can modify the energy spectrum in a nonhomogeneous way. Therefore, the Rashba interaction acts as a quantum control parameter capable of changing both the magnitude of the energy exchanges and the operation mode of the cycle.

In what follows we adopt the sign convention introduced in the previous section, in which positive work means work extracted from the working substance. Thus, the operating regimes are classified according to the signs of $Q_h$, $Q_c$, and $W=Q_h+Q_c$, as summarized in Table~\ref{tab:regimes}. In the heat-engine regime the working substance absorbs heat from the hot reservoir, releases heat to the cold reservoir and produces useful work. In the refrigerator regime the cycle consumes work in order to extract heat from the cold reservoir. The heater and accelerator regimes correspond to dissipative modes in which work is supplied to the system without refrigeration.

\begin{table}[h]
\caption[Thermodynamic regimes]{Thermodynamic regimes of operation according to the signs of the heat exchanges and total work.}
\label{tab:regimes}
\begin{ruledtabular}
\begin{tabular}{cccc}
Regime & $Q_h$ & $Q_c$ & $W$ \tabularnewline
\hline
Heat engine & $+$ & $-$ & $+$ \tabularnewline
Refrigerator & $-$ & $+$ & $-$ \tabularnewline
Heater & $-$ & $-$ & $-$ \tabularnewline
Accelerator & $+$ & $-$ & $-$ \tabularnewline
\end{tabular}
\end{ruledtabular}
\end{table}

\begin{figure*}[!t]
\centering
\begin{minipage}{0.68\textwidth}
\centering
\textbf{(a)}
\includegraphics[width=\linewidth]{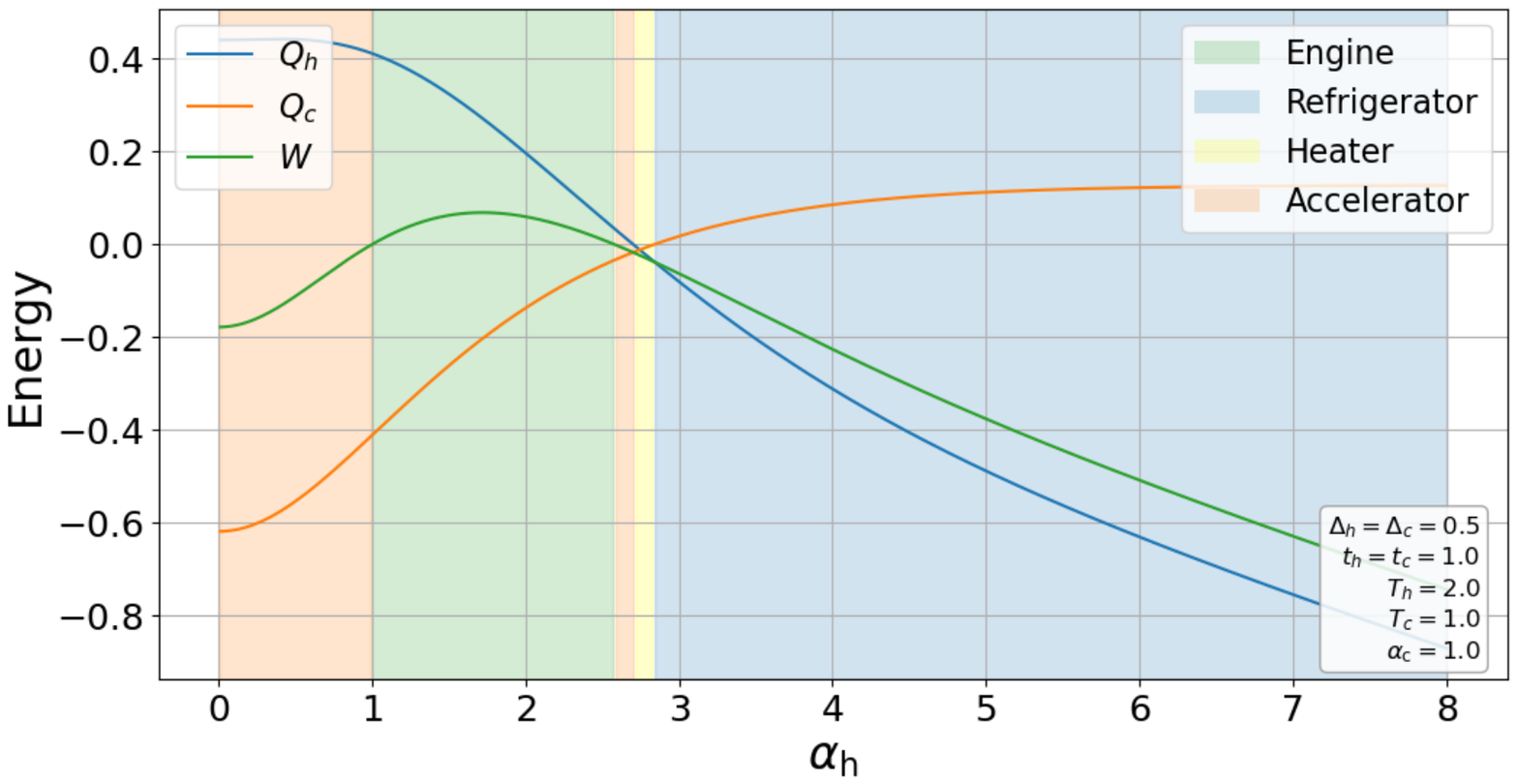}
\end{minipage}

\vspace{0.25cm}

\begin{minipage}{0.68\textwidth}
\centering
\textbf{(b)}
\includegraphics[width=\linewidth]{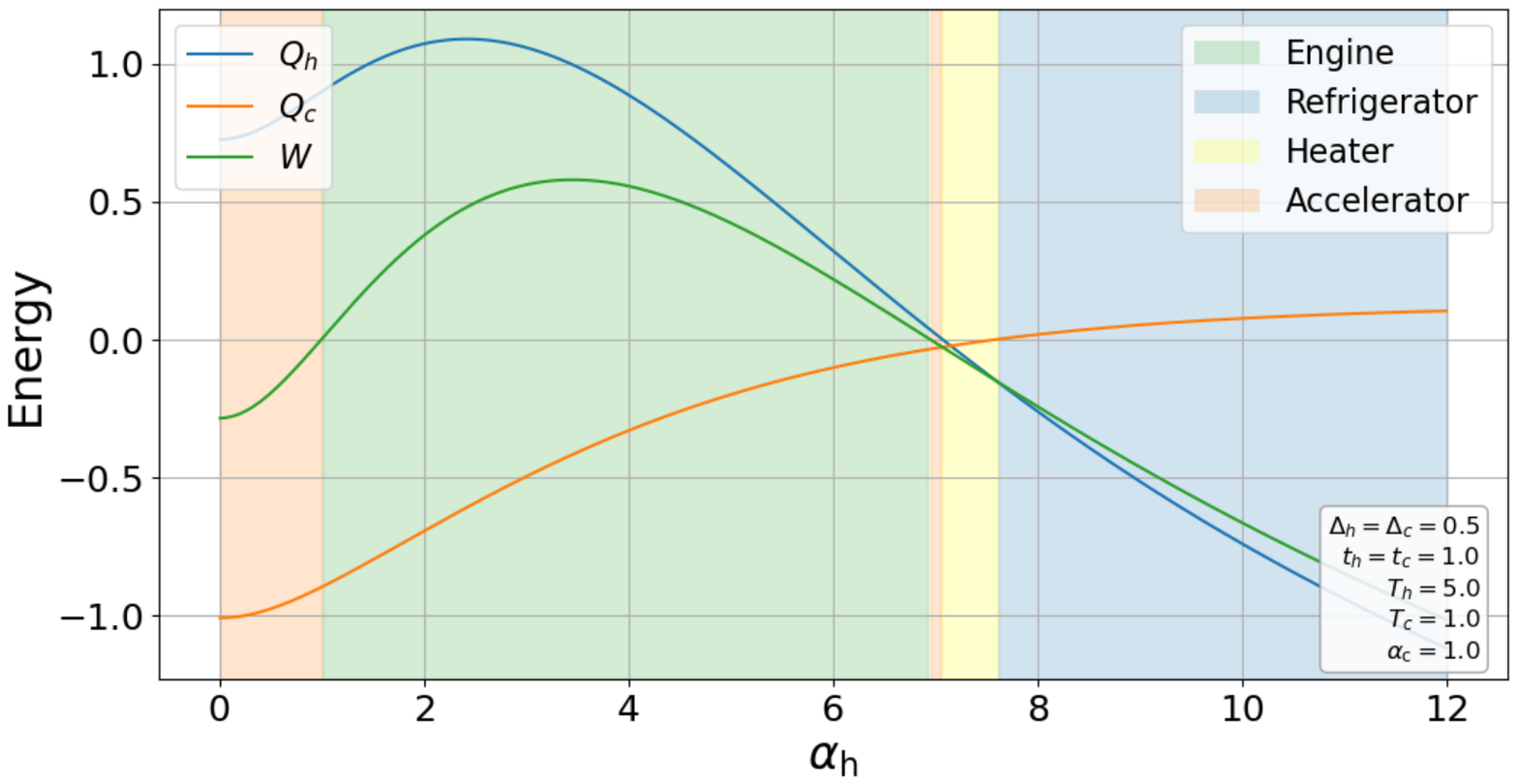}
\end{minipage}
\caption[Rashba scan of heat and work]{Heat exchanges $Q_h$ and $Q_c$, and total work $W$, as functions of the Rashba coupling $\alpha_h$. The fixed parameters are $\Delta_h=\Delta_c=0.5$, $t_h=t_c=1.0$, and $\alpha_c=1.0$. Panel (a) corresponds to $T_h=2.0 $ and $T_c=1.0 $, while panel (b) corresponds to $T_h=5.0 $ and $T_c=1.0 $. The colored regions indicate the thermodynamic regimes defined in Table~\ref{tab:regimes}.}
\label{fig:var-alpha-h}
\end{figure*}

We first consider one-dimensional scans in which the Rashba coupling during the hot isochoric stroke, $\alpha_h$, is varied while the remaining parameters are kept fixed. In Fig.~\ref{fig:var-alpha-h} we show $Q_h$, $Q_c$, and $W$ as functions of $\alpha_h$ for two different pairs of reservoir temperatures, namely $(T_h,T_c)=(2.0  ,1.0  )$ and $(T_h,T_c)=(5.0  ,1.0  )$. In both cases we set $\Delta_h=\Delta_c=0.5$, $t_h=t_c=1.0$, and $\alpha_c=1.0$. The colored background identifies the thermodynamic regime at each value of $\alpha_h$.

The main effect observed in Fig.~\ref{fig:var-alpha-h}  concerning heat-engine regime is that increasing the temperature difference enlarges its corresponding window and increases the maximum work output. For $T_h=2.0  $ and $T_c=1.0  $, the engine regime occurs only in a narrow interval around $\alpha_{h}\approx 0.5$ and $\alpha_{h}\approx 2.5$, whereas for $T_h=5.0 $ and $T_c=1 $ this interval becomes much broader, from $\alpha_{h}\approx 0.5$ until $\alpha_{h}\approx 7.0$. Thus, the same variation of the Rashba coupling may produce different operational regimes depending on the temperature gradient imposed by the two reservoirs.

\begin{figure*}[!t]
\centering
\includegraphics[width=\textwidth]{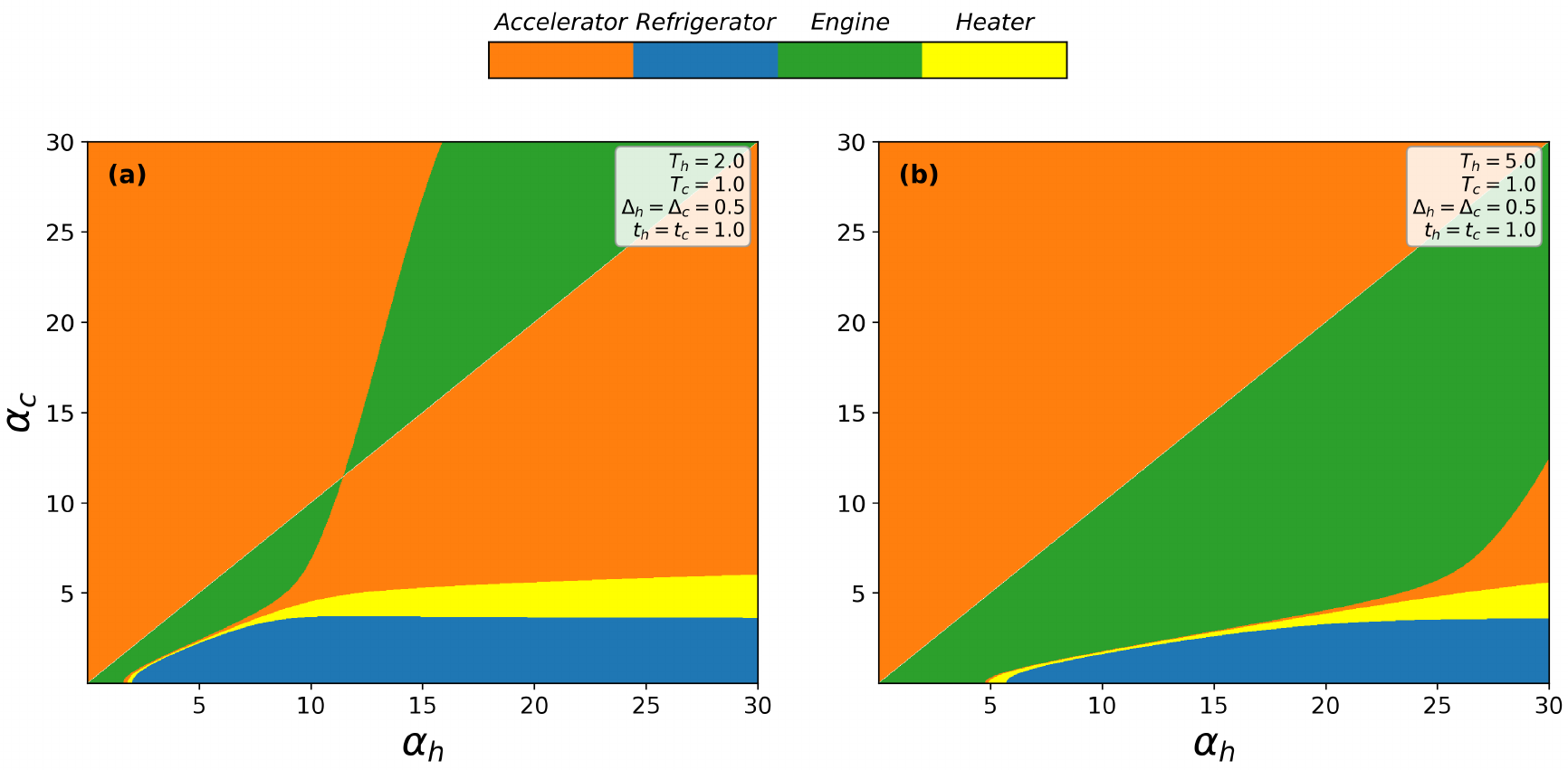}
\caption[Thermodynamic-regime maps]{Thermodynamic-regime maps in the $(\alpha_h,\alpha_c)$ plane for fixed $\Delta_h=\Delta_c=0.5$ and $t_h=t_c=1.0$. Panel (a) corresponds to $T_h=2.0$ and $T_c=1.0$, while panel (b) corresponds to $T_h=5.0$ and $T_c=1.0$. The colors indicate accelerator, refrigerator, heat-engine, and heater regimes. The scans in Fig.~\ref{fig:var-alpha-h} are horizontal cuts of these diagrams at $\alpha_c=1.0$.}
\label{fig:regime-maps}
\end{figure*}

The previous plots correspond to horizontal cuts, at fixed $\alpha_c=1.0$, of the complete regime maps in the $(\alpha_h,\alpha_c)$ plane. These maps are shown in Fig.~\ref{fig:regime-maps}. They make clear that the thermodynamic behavior is governed by the relative change of the Rashba coupling between the hot and cold configurations. The diagonal structure separating broad regions of the plots indicates that the machine is particularly sensitive to the mismatch between $\alpha_h$ and $\alpha_c$.

As shown in Fig.~\ref{fig:regime-maps}(a) and Fig.~\ref{fig:regime-maps}(b), when the temperature difference between the reservoirs increases, the regime map evolves toward a simpler diagonal structure. The lower diagonal sector, where $\alpha_h$ is larger than $\alpha_c$ in the appropriate range, is progressively converted into heat-engine operation. In contrast, the upper diagonal sector, associated with comparatively larger $\alpha_c$, is dominated by the accelerator regime. This behavior shows that increasing the temperature gradient imposed by the two reservoirs stabilizes work extraction when the Rashba coupling is larger in the hot configuration, while the opposite imbalance favors a dissipative mode in which the system still absorbs heat from the hot reservoir but requires external work. The refrigerator and heater regions are then reduced to narrower boundary domains.

We now analyze the performance of the operational regimes. In the heat-engine regime, the relevant quantity is the efficiency \cite{myers-1}
\begin{eqnarray}
\eta
&=&
\frac{W}{Q_h},
\label{ResultsEta}
\end{eqnarray}
with $Q_h>0$ and $W>0$. For the remaining regimes, we first define the regime-dependent performance coefficients according to the useful heat flow in each operation mode \cite{moi-1}:
\begin{eqnarray}
\tilde{\kappa}_{\mathrm{ref}}
&=&
\frac{Q_h}{W}, \\
\tilde{\kappa}_{\mathrm{heat}}
&=&
\frac{Q_c}{W}, \\
\tilde{\kappa}_{\mathrm{acc}}
&=&
\frac{Q_c}{W},
\label{ResultsKappa}
\end{eqnarray}
These ratios are positive in their corresponding regimes because both numerator and denominator have the same sign. Thus, they should be understood as regime-dependent performance coefficients $\tilde{\kappa}$ rather than a single universal quantity. For visualization purposes, each coefficient $\tilde{\kappa}$ is mapped to the bounded quantity \cite{moi-1}
\begin{eqnarray}
\kappa
&=&
\frac{\tilde{\kappa}}{1+\tilde{\kappa}},
\label{ResultsKappaBounded}
\end{eqnarray}
so that both $\eta$ and the non-engine performance coefficient $\kappa$ can be displayed on the same color scale between zero and one.

\begin{figure*}[!t]
\centering
\includegraphics[width=\textwidth]{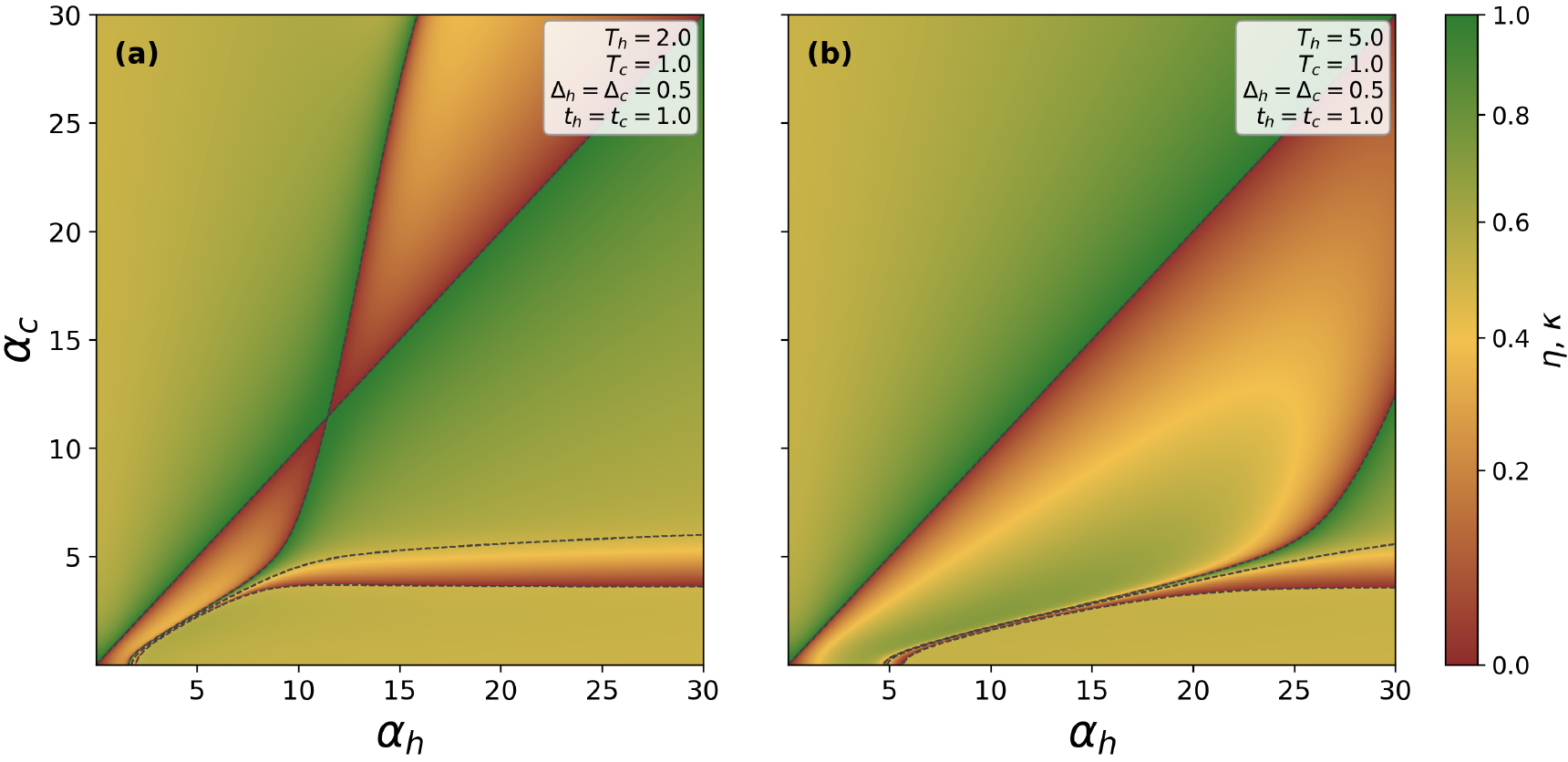}
\caption[Efficiency and performance map]{Efficiency $\eta$ and bounded regime-dependent performance coefficient $\kappa$ in the $(\alpha_h,\alpha_c)$ plane for fixed $\Delta_h=\Delta_c=0.5$ and $t_h=t_c=1.0$. Panel (a) corresponds to $T_h=2.0$ and $T_c=1.0$, while panel (b) corresponds to $T_h=5.0$ and $T_c=1.0$. In the heat-engine region the color represents $\eta$, whereas in the refrigerator, heater, and accelerator regions it represents $\kappa=\tilde{\kappa}/(1+\tilde{\kappa})$, with $\tilde{\kappa}$ defined according to Eq.~(\ref{ResultsKappa}). The color scale therefore measures the performance of the active thermodynamic regime, with darker red tones corresponding to low performance and green tones corresponding to values close to unity.}
\label{fig:eta-kappa-map}
\end{figure*}

Figure~\ref{fig:eta-kappa-map} combines the efficiency of the engine and the bounded performance coefficient $\kappa$ of the non-engine regimes for the same temperature pairs used in the regime maps of Fig.~\ref{fig:regime-maps}. The color scale encodes the performance of the regime realized at each point of the $(\alpha_h,\alpha_c)$ plane: in the engine region it represents $\eta$, while in the refrigerator, heater, and accelerator regions it represents $\kappa$. A particularly relevant feature appears close to the transition between the heat-engine and accelerator regimes, especially in the larger-temperature-gradient case shown in panel (b). On the accelerator side of this boundary, the coefficient $\kappa$ can be high, whereas on the heat-engine side the efficiency reaches values next to zero. Thus, crossing the boundary into the engine regime is not sufficient to obtain an efficient thermal machine: near the transition, the signs of $Q_h$, $Q_c$, and $W$ already correspond to work extraction, but the extracted work remains small compared with the heat absorbed from the hot reservoir. Higher efficiencies are found in a nontrivial region inside the engine domain, away from the immediate transition line. 

\begin{figure*}[!t]
\centering
\includegraphics[width=0.55\textwidth]{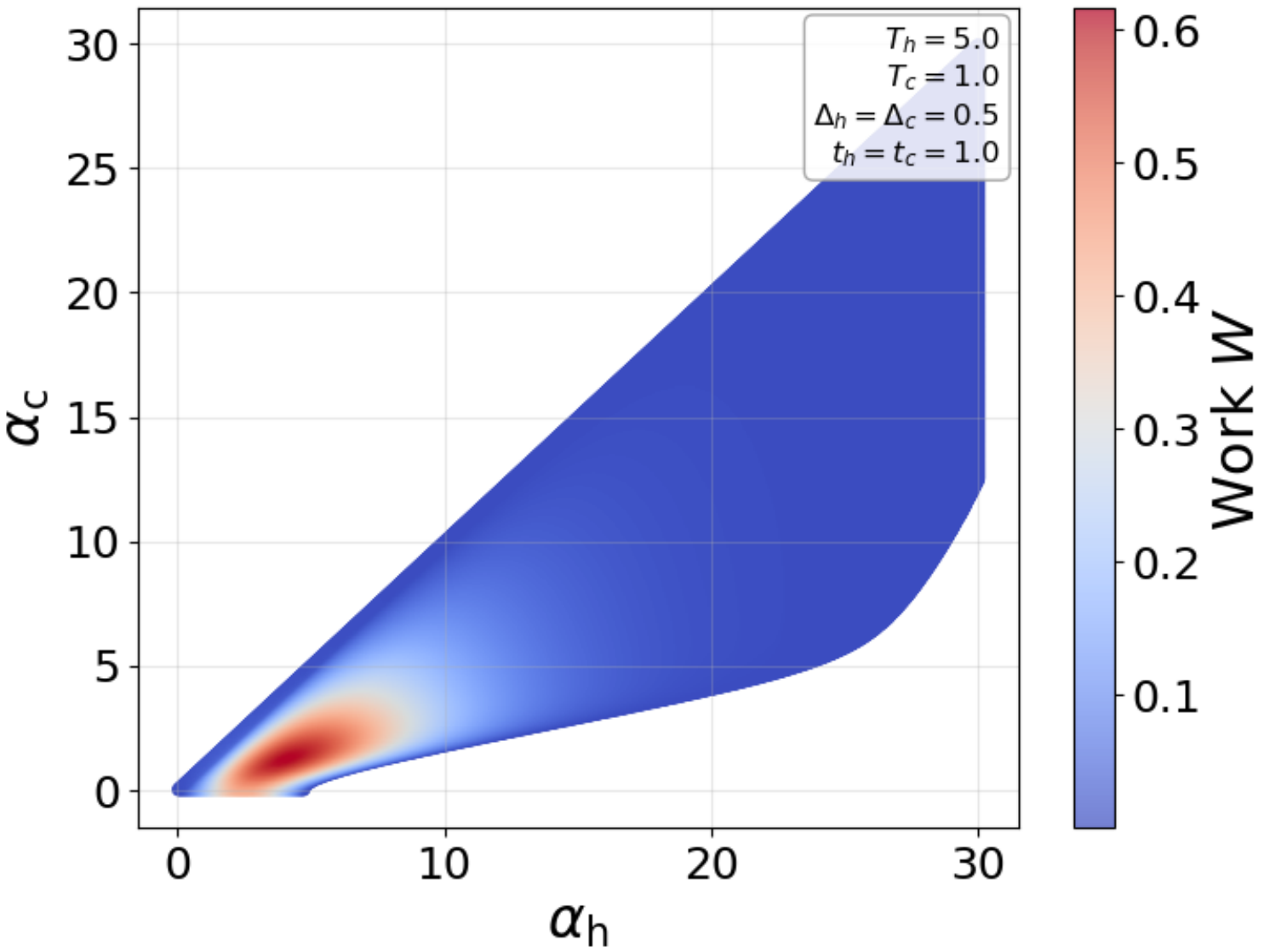}
\caption[Work output in the heat-engine regime]{Work output $W$ in the heat-engine regime in the $(\alpha_h,\alpha_c)$ plane for $T_h=5.0$, $T_c=1.0$, $\Delta_h=\Delta_c=0.5$, and $t_h=t_c=1.0$. The color scale represents the extracted work. The maximum-work region is localized close to the lower boundary of the diagram, where $\alpha_c$ is very small.}
\label{fig:engine-work-map}
\end{figure*}

We now restrict the discussion to the heat-engine regime and focus first on the work output. Figure~\ref{fig:engine-work-map} shows the values of $W$ inside the engine region in the $(\alpha_h,\alpha_c)$ plane for $T_h=5.0  $, $T_c=1.0  $, $\Delta_h=\Delta_c=0.5$, and $t_h=t_c=1.0$. In contrast with the broad heat-engine domain seen in Fig.~\ref{fig:regime-maps}, appreciable work is concentrated in a localized region where $\alpha_c$  is around $0.0$ and $2.5$, while $\alpha_{h}$ is around $2.5$ and $6.0$. Comparing this map with Fig.~\ref{fig:eta-kappa-map}, we see that this high-work region partially overlaps with the inner engine region where the efficiency is also high. Thus, the most relevant operating window is not simply the whole heat-engine phase, but the subset in which sizable work output and high efficiency coexist.

Motivated by this observation, we performed a global numerical search for the maximum values of the efficiency and the work output in the heat-engine regime. In this calculation all Hamiltonian and temperature parameters were varied, with the restrictions $T_h>T_c$, $T_h,T_c\in[1.0 ,5.0 ]$, $\Delta_h=\Delta_c\equiv\Delta\in[0.0,5.0]$, and $t_h=t_c\equiv t\in[0.0,5.0]$. The Rashba couplings $\alpha_h$ and $\alpha_c$ were varied independently in the interval $[0.0,20.0]$. The calculation was carried out in the Curie cluster at UFLA, using the SLURM batch system. The production run used a single-node batch job, Python 3.11.6, and a memory allocation of 16 GB. A total of $184\,480\,000$ parameter points were analyzed, among which $57\,200\,729$ points satisfied the heat-engine conditions.

\begin{table*}[!t]
\caption[Global maxima in the engine regime]{Top fifteen global maxima of efficiency and work in the heat-engine regime. All listed points have $T_h=5.0$, $T_c=1.0$, and $\alpha_c=0.0$.}
\label{tab:global-maxima}
\scriptsize
\begin{ruledtabular}
\begin{tabular}{cccccccc|cccccccc}
\multicolumn{8}{c|}{Largest $\eta$} & \multicolumn{8}{c}{Largest $W$} \tabularnewline
$n$ & $\eta$ & $W$ & $\Delta$ & $t$ & $\alpha_h$ & $T_h$ & $T_c$ &
$n$ & $\eta$ & $W$ & $\Delta$ & $t$ & $\alpha_h$ & $T_h$ & $T_c$ \tabularnewline
\hline
$1$ & $0.799696$ & $0.001261$ & $0.0$ & $0.917719$ & $4.492191$ & $5.0$ & $1.0$ & $1$ & $0.576557$ & $0.626560$ & $0.0$ & $1.796764$ & $3.844035$ & $5.0$ & $1.0$ \tabularnewline
$2$ & $0.799695$ & $0.001113$ & $0.0$ & $0.917719$ & $4.492625$ & $5.0$ & $1.0$ & $2$ & $0.576596$ & $0.626560$ & $0.0$ & $1.796764$ & $3.844469$ & $5.0$ & $1.0$ \tabularnewline
$3$ & $0.799694$ & $0.001147$ & $0.0$ & $0.905161$ & $4.431020$ & $5.0$ & $1.0$ & $3$ & $0.576518$ & $0.626560$ & $0.0$ & $1.796764$ & $3.843601$ & $5.0$ & $1.0$ \tabularnewline
$4$ & $0.799694$ & $0.001296$ & $0.0$ & $0.905161$ & $4.430586$ & $5.0$ & $1.0$ & $4$ & $0.576635$ & $0.626560$ & $0.0$ & $1.796764$ & $3.844902$ & $5.0$ & $1.0$ \tabularnewline
$5$ & $0.799693$ & $0.001181$ & $0.0$ & $0.892603$ & $4.369414$ & $5.0$ & $1.0$ & $5$ & $0.576478$ & $0.626560$ & $0.0$ & $1.796764$ & $3.843167$ & $5.0$ & $1.0$ \tabularnewline
$6$ & $0.799693$ & $0.001409$ & $0.0$ & $0.917719$ & $4.491757$ & $5.0$ & $1.0$ & $6$ & $0.576674$ & $0.626560$ & $0.0$ & $1.796764$ & $3.845336$ & $5.0$ & $1.0$ \tabularnewline
$7$ & $0.799692$ & $0.001330$ & $0.0$ & $0.892603$ & $4.368980$ & $5.0$ & $1.0$ & $7$ & $0.576439$ & $0.626560$ & $0.0$ & $1.796764$ & $3.842733$ & $5.0$ & $1.0$ \tabularnewline
$8$ & $0.799692$ & $0.001215$ & $0.0$ & $0.880045$ & $4.307809$ & $5.0$ & $1.0$ & $8$ & $0.576714$ & $0.626560$ & $0.0$ & $1.796764$ & $3.845770$ & $5.0$ & $1.0$ \tabularnewline
$9$ & $0.799690$ & $0.001444$ & $0.0$ & $0.905161$ & $4.430152$ & $5.0$ & $1.0$ & $9$ & $0.576400$ & $0.626560$ & $0.0$ & $1.796764$ & $3.842299$ & $5.0$ & $1.0$ \tabularnewline
$10$ & $0.799690$ & $0.001032$ & $0.0$ & $0.892603$ & $4.369848$ & $5.0$ & $1.0$ & $10$ & $0.576584$ & $0.626560$ & $3.593246$ & $0.0$ & $3.844035$ & $5.0$ & $1.0$ \tabularnewline
$11$ & $0.799690$ & $0.001065$ & $0.0$ & $0.880045$ & $4.308243$ & $5.0$ & $1.0$ & $11$ & $0.576544$ & $0.626560$ & $3.593246$ & $0.0$ & $3.843601$ & $5.0$ & $1.0$ \tabularnewline
$12$ & $0.799690$ & $0.001248$ & $0.0$ & $0.867487$ & $4.246204$ & $5.0$ & $1.0$ & $12$ & $0.576753$ & $0.626560$ & $0.0$ & $1.796764$ & $3.846204$ & $5.0$ & $1.0$ \tabularnewline
$13$ & $0.799689$ & $0.000999$ & $0.0$ & $0.905161$ & $4.431453$ & $5.0$ & $1.0$ & $13$ & $0.576623$ & $0.626560$ & $3.593246$ & $0.0$ & $3.844469$ & $5.0$ & $1.0$ \tabularnewline
$14$ & $0.799689$ & $0.001098$ & $0.0$ & $0.867487$ & $4.246638$ & $5.0$ & $1.0$ & $14$ & $0.576505$ & $0.626560$ & $3.593246$ & $0.0$ & $3.843167$ & $5.0$ & $1.0$ \tabularnewline
$15$ & $0.799689$ & $0.001364$ & $0.0$ & $0.880045$ & $4.307375$ & $5.0$ & $1.0$ & $15$ & $0.576662$ & $0.626560$ & $3.593246$ & $0.0$ & $3.844902$ & $5.0$ & $1.0$ \tabularnewline
\end{tabular}
\end{ruledtabular}
\end{table*}

The fifteen largest global values of $\eta$ and $W$ found in the heat-engine regime are shown in Table~\ref{tab:global-maxima}. The highest efficiencies are obtained for the largest temperature gradient imposed by the two reservoirs in the scan, $T_h=5.0 $ and $T_c=1.0 $, with $\alpha_c=\Delta = 0.0$. The maximum efficiency is close to  $\eta\simeq0.80$, which is also the Carnot efficiency for $T_h=5.0 $ and $T_c=1.0 $. However, the work associated with these high-efficiency points is very small, of order $10^{-3}$. This shows that the best efficiency is reached close to a weak-output boundary of the engine domain.

The global maximum of work is $W\simeq0.63$, around three orders of magnitude larger than the work obtained at the maximum-efficiency points. In this case the efficiency remains appreciable, $\eta\simeq0.58$, although it is significantly below the global efficiency maximum. Again, the optimal points occur at $T_h=5.0 $, $T_c=1.0 $, and $\alpha_c =0.0$. The initial maximum-work point occur in $\Delta=0.0$, with the tunneling parameter assuming a finite value of $t\simeq1.8$, and the Rashba coupling at $\alpha_h\simeq3.8$. However, the extended ranking also displays nearly degenerate maxima in which this structure is interchanged: $t$ becomes very small while $\Delta$ assumes a finite value close to $3.6$, with $\alpha_h$ remaining in the same region around $3.8$. This alternation between $\Delta$ and $t$ follows from the structure of the energy gaps in Eqs.~(\ref{Model8}) and (\ref{Model9}), which contain terms of the form $\sqrt{(\Delta\pm2t)^2+4\alpha^2}$. The Zeeman splitting and the tunneling amplitude therefore enter the spectrum through the same combinations $\Delta\pm2t$, while the Rashba coupling contributes through $4\alpha^2$. Since the work of the Otto cycle depends on the change of these gaps between the hot and cold configurations, the maximum work is obtained from an optimal spectral deformation, not from independently maximizing any single Hamiltonian parameter. In this sense, the persistence of $\alpha_h\simeq3.8$ across the nearly degenerate maxima indicates that the Rashba coupling fixes the optimal spin-orbit scale, while $\Delta$ and $t$ can partially compensate each other through the gap combinations.

Since $\Delta=0.0$ corresponds to the case of absence of Zeeman splitting, it is useful to compare these results with a scan restricted to a nonzero splitting. Table~\ref{tab:delta-one-maxima} shows the ten largest values of $\eta$ and $W$ among the engine points with $\Delta$ close to one. In this subset, $288\,495$ engine points were found. The best efficiencies decrease to $\eta\simeq0.46$, but the work is now much larger than in the global efficiency ranking, $W\simeq0.43$. The maximum-work points for $\Delta\simeq1$ reach $W\simeq0.51$, with efficiencies around $\eta\simeq0.42$. Thus, fixing a finite Zeeman splitting reduces the maximum attainable performance when compared with the $\Delta=0.0$ global optimum, but it preserves a robust heat-engine regime with sizable work output.

\begin{table*}[!t]
\caption[Maxima near nonzero Zeeman splitting]{Top ten maxima of efficiency and work in the heat-engine regime for $\Delta$ close to one. All listed points have $T_h=5.0$, $T_c=1.0$, $\alpha_c=0.0$, and $\Delta=1.0$ in the numerical grid.}
\label{tab:delta-one-maxima}
\scriptsize
\begin{ruledtabular}
\begin{tabular}{cccccccc|cccccccc}
\multicolumn{8}{c|}{Largest $\eta$ for $\Delta\simeq1$} & \multicolumn{8}{c}{Largest $W$ for $\Delta\simeq1$} \tabularnewline
$n$ & $\eta$ & $W$ & $\Delta$ & $t$ & $\alpha_h$ & $T_h$ & $T_c$ &
$n$ & $\eta$ & $W$ & $\Delta$ & $t$ & $\alpha_h$ & $T_h$ & $T_c$ \tabularnewline
\hline
$1$ & $0.458327$ & $0.429163$ & $1.0$ & $1.570724$ & $4.368113$ & $5.0$ & $1.0$ & $1$ & $0.415095$ & $0.506973$ & $1.0$ & $1.846995$ & $3.672668$ & $5.0$ & $1.0$ \tabularnewline
$2$ & $0.458327$ & $0.429114$ & $1.0$ & $1.570724$ & $4.368547$ & $5.0$ & $1.0$ & $2$ & $0.415120$ & $0.506973$ & $1.0$ & $1.846995$ & $3.673102$ & $5.0$ & $1.0$ \tabularnewline
$3$ & $0.458327$ & $0.429211$ & $1.0$ & $1.570724$ & $4.367679$ & $5.0$ & $1.0$ & $3$ & $0.415069$ & $0.506973$ & $1.0$ & $1.846995$ & $3.672234$ & $5.0$ & $1.0$ \tabularnewline
$4$ & $0.458327$ & $0.429065$ & $1.0$ & $1.570724$ & $4.368980$ & $5.0$ & $1.0$ & $4$ & $0.415145$ & $0.506973$ & $1.0$ & $1.846995$ & $3.673536$ & $5.0$ & $1.0$ \tabularnewline
$5$ & $0.458327$ & $0.429260$ & $1.0$ & $1.570724$ & $4.367245$ & $5.0$ & $1.0$ & $5$ & $0.415044$ & $0.506973$ & $1.0$ & $1.846995$ & $3.671800$ & $5.0$ & $1.0$ \tabularnewline
$6$ & $0.458327$ & $0.429016$ & $1.0$ & $1.570724$ & $4.369414$ & $5.0$ & $1.0$ & $6$ & $0.415171$ & $0.506973$ & $1.0$ & $1.846995$ & $3.673970$ & $5.0$ & $1.0$ \tabularnewline
$7$ & $0.458327$ & $0.429309$ & $1.0$ & $1.570724$ & $4.366811$ & $5.0$ & $1.0$ & $7$ & $0.415018$ & $0.506973$ & $1.0$ & $1.846995$ & $3.671367$ & $5.0$ & $1.0$ \tabularnewline
$8$ & $0.458327$ & $0.428967$ & $1.0$ & $1.570724$ & $4.369848$ & $5.0$ & $1.0$ & $8$ & $0.415196$ & $0.506973$ & $1.0$ & $1.846995$ & $3.674403$ & $5.0$ & $1.0$ \tabularnewline
$9$ & $0.458327$ & $0.429358$ & $1.0$ & $1.570724$ & $4.366377$ & $5.0$ & $1.0$ & $9$ & $0.414993$ & $0.506973$ & $1.0$ & $1.846995$ & $3.670933$ & $5.0$ & $1.0$ \tabularnewline
$10$ & $0.458327$ & $0.428918$ & $1.0$ & $1.570724$ & $4.370282$ & $5.0$ & $1.0$ & $10$ & $0.415222$ & $0.506973$ & $1.0$ & $1.846995$ & $3.674837$ & $5.0$ & $1.0$ \tabularnewline
\end{tabular}
\end{ruledtabular}
\end{table*}

Finally, Fig.~5 summarizes how the engine points populate the $(W,\eta)$ plane when the hot-reservoir temperature is separated into intervals. Each panel contains heat-engine points from the global search, grouped according to the range of $T_h$, while the color indicates the cold-reservoir temperature $T_c$. The accessible region grows as $T_h$ increases. For $T_h\in[1.0,2.0]$, the work output is small and the efficiency remains below approximately $0.5$. As the hot reservoir becomes hotter, both the maximum efficiency and the maximum work increase, and the point cloud spreads toward larger $W$. The panel with $T_h\in[4.0,5.0]$ contains the global maxima discussed above and shows the widest engine domain in the $(W,\eta)$ plane. In other words, the temperature difference mainly sets the maximum efficiency that can be reached, as expected from the thermodynamic bound associated with the reservoirs, while the Hamiltonian parameters, especially the Rashba couplings and the tunneling amplitude, determine how much work can actually be extracted from this available temperature gradient imposed by the two reservoirs. The color distribution also shows that the best performance is associated with the smallest available $T_c$, consistently with the appearance of $T_c=1.0$ in all the top-ranked points.

\onecolumngrid
\begin{center}
\centering
\includegraphics[width=0.82\textwidth]{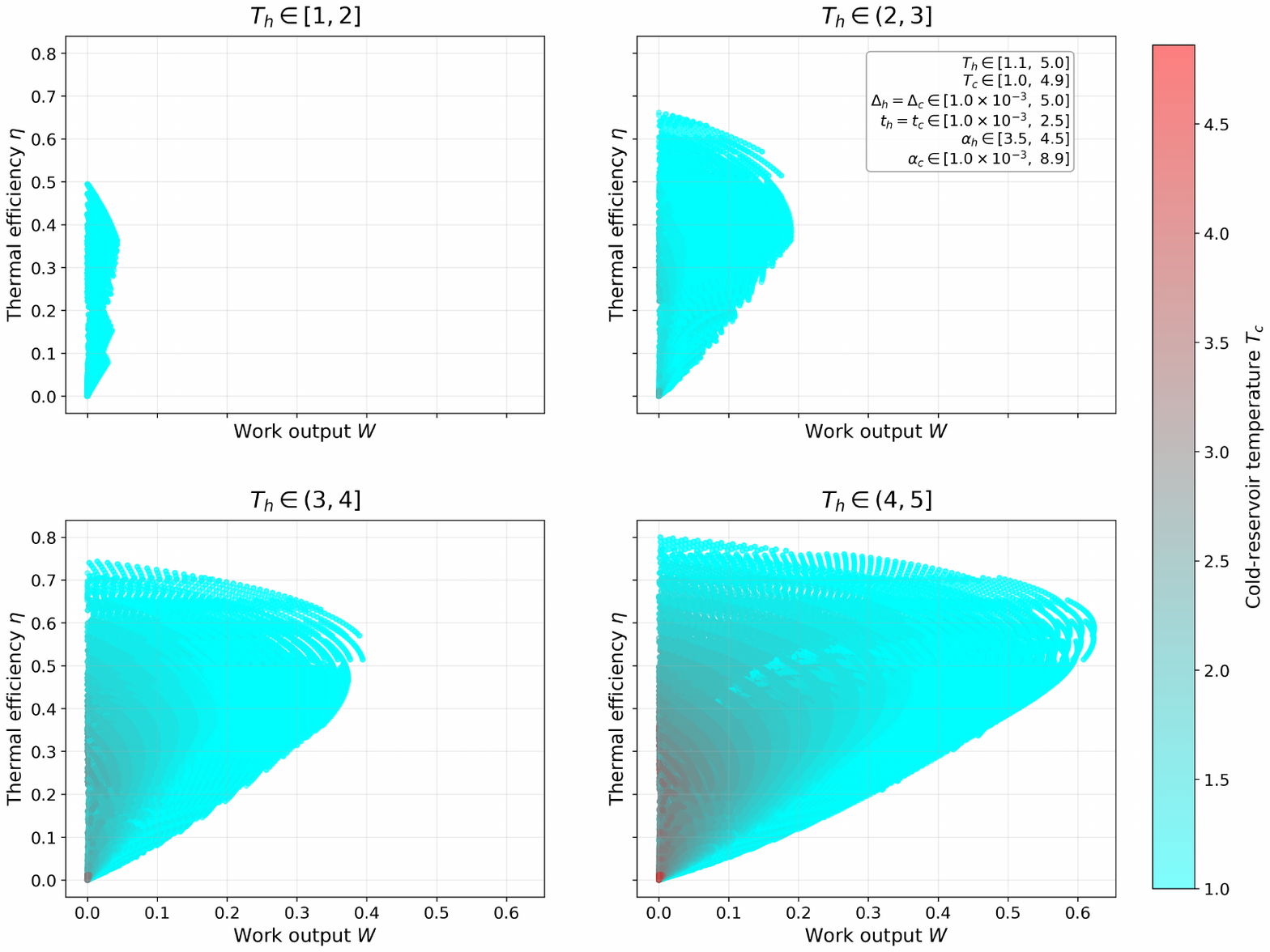}
\par\smallskip
\begin{minipage}{0.82\textwidth}
\small
\textbf{Figure 5.} Heat-engine points in the $(W,\eta)$ plane grouped by hot-reservoir temperature intervals. Each panel corresponds to a range of $T_h$, while the color scale represents $T_c$. The plot shows how the accessible efficiency-work domain expands as the temperature gradient imposed by the two reservoirs is increased.
\end{minipage}
\end{center}
\twocolumngrid

\FloatBarrier

\section{Conclusions}
\label{conclusionsAndPerspectives}

In this work we investigated the thermodynamic operation of a quantum Otto machine whose working substance is a single electron confined in a double quantum dot with Rashba spin-orbit interaction. The Hamiltonian parameters $\Delta$, $t$, and $\alpha$ were treated as externally controllable quantities. Within this broader parameter space, we focused on the role of the Rashba interaction.

The regime maps in the $(\alpha_h,\alpha_c)$ plane show that the Rashba interaction is an efficient control knob for switching between heat-engine, refrigerator, heater, and accelerator regimes. As the temperature gradient imposed by the two reservoirs increases, the lower diagonal sector of the map is progressively converted into heat-engine operation, while the upper diagonal sector becomes dominated by the accelerator regime. We then focused on the heat-engine regime to analyze efficiency and work output in more detail. This analysis showed that there is a small window where relatively high efficiency and work output coexist.

To identify this operating window, we performed a robust numerical search in the Curie cluster, varying the Hamiltonian parameters and reservoir temperatures over a large grid and classifying the points according to their thermodynamic regime. The search revealed that the largest efficiencies are obtained for the largest temperature gradient considered and for $\Delta$ approaching zero, reaching values close to the Carnot bound but with very small work output. In contrast, the maximum-work points occur at finite tunneling and finite Rashba coupling, with lower but still significant efficiency. When the Zeeman splitting is constrained to remain finite, $\Delta=1.0$, the maximum efficiency decreases, but the work output remains sizable, indicating a more favorable compromise between efficiency and useful work production. Overall, the results show that the performance of this quantum thermal machine is governed by the combined spectral deformation produced by $\Delta$, $t$, and $\alpha$, rather than by the temperature gradient alone.

\begin{acknowledgments}
M. Rojas acknowledge the financial support from the CNPq and FAPEMIG.
M. Rojas also gratefully acknowledges CNPq for support through Grant No. 311565/2025-5. 
\end{acknowledgments}

\end{document}